\documentclass[%
 aip,
 amsmath,amssymb,
 reprint,%
]{revtex4-1}

\usepackage{graphicx}
\usepackage{dcolumn}
\usepackage{bm}

\usepackage[utf8]{inputenc}
\usepackage[T1]{fontenc}
\usepackage{mathptmx}
\usepackage{etoolbox}
\usepackage{xcolor}
\usepackage{SIunits}
\usepackage[textsize=tiny]{todonotes}

\usepackage{caption}

\makeatletter
\def\@email#1#2{%
 \endgroup
 \patchcmd{\titleblock@produce}
  {\frontmatter@RRAPformat}
  {\frontmatter@RRAPformat{\produce@RRAP{*#1\href{mailto:#2}{#2}}}\frontmatter@RRAPformat}
  {}{}
}%
\makeatother
\begin{document}

\newcommand{\fg}{\textcolor{red} }
\newcommand{\tf}{\textcolor{blue} }
\newcommand{\vr}{\textcolor{teal} }
\newcommand{\lb}{\textcolor{orange} }

\newcommand{\luigi}[2][]{\todo[color=orange!20,#1]{{\bf LB:} #2}}
\newcommand{\valerio}[2][]{\todo[color=teal!20,#1]{{\bf VR:} #2}}
\newcommand{\thorben}[2][]{\todo[color=blue!20,#1]{{\bf TF:} #2}}
\newcommand{\francesco}[2][]{\todo[color=red!20,#1]{{\bf FG:} #2}}

\preprint{AIP/123-QED}

\title{Deep learning path-like collective variable for enhanced sampling molecular dynamics}

\author{Thorben Fr\"ohlking}%
\affiliation{School of Pharmaceutical Sciences, University of Geneva, Rue Michel Servet 1, 1206, Genève, Switzerland}
\affiliation{Institute of Pharmaceutical Sciences of Western Switzerland (ISPSO), University of Geneva, 1206, Genève, Switzerland}
\affiliation{Swiss Institute of Bioinformatics, University of Geneva, 1206, Genève, Switzerland}

\author{Luigi Bonati}%
\affiliation{Italian Institute of Technology, Via Melen 83, 16152 Genoa, Italy}

\author{Valerio Rizzi}%
\affiliation{School of Pharmaceutical Sciences, University of Geneva, Rue Michel Servet 1, 1206, Genève, Switzerland}
\affiliation{Institute of Pharmaceutical Sciences of Western Switzerland (ISPSO), University of Geneva, 1206, Genève, Switzerland}
\affiliation{Swiss Institute of Bioinformatics, University of Geneva, 1206, Genève, Switzerland}

\author{Francesco Luigi Gervasio}
\email{francesco.gervasio@unige.ch}
\affiliation{School of Pharmaceutical Sciences, University of Geneva, Rue Michel Servet 1, 1206, Genève, Switzerland}
\affiliation{Institute of Pharmaceutical Sciences of Western Switzerland (ISPSO), University of Geneva, 1206, Genève, Switzerland}
\affiliation{Swiss Institute of Bioinformatics, University of Geneva, 1206, Genève, Switzerland}
\affiliation{Department of Chemistry, University College London, London, WC1E 6BT, United Kingdom}


\date{\today}

\begin{abstract}


Several enhanced sampling techniques rely on the definition of collective variables to effectively explore free energy landscapes. Existing variables that describe the progression along a reactive pathway offer an elegant solution but face a number of limitations. In this paper, we address these challenges by introducing a new path-like collective variable called the `Deep-locally-non-linear-embedding', which is inspired by principles of the locally linear embedding technique and is trained on a reactive trajectory. The variable mimics the ideal reaction coordinate by automatically generating a non-linear combination of features through a differentiable generalized autoencoder that combines a neural network with a continuous k-nearest-neighbor selection. Among the key advantages of this method is its capability to automatically choose the metric for searching neighbors and to learn the path from state A to state B without the need to handpick landmarks \textit{a priori}. We demonstrate the effectiveness of DeepLNE by showing that the progression along the path variable closely approximates the ideal reaction coordinate in toy models such as the M\"uller-Brown-potential and alanine dipeptide. We then use it in molecular dynamics simulations of an RNA tetraloop, where we highlight its capability to accelerate transitions and converge the free energy of folding.

\end{abstract}

\maketitle

Atomistic molecular dynamics (MD) simulations have proven to be a powerful tool for investigating intricate aspects of physical, chemical and biological systems.~\cite{Frenkel} The power of MD investigations has increased with advances in computational resources and refinement of force field accuracy. Still, many phenomena of importance exhibit timescales well beyond the reach of conventional unbiased MD, even when used on state-of-the-art supercomputing platforms. In response to this temporal constraint, a variety of \textit{enhanced sampling} algorithms have been developed over time to augment the sampling methods and reconstruct the corresponding free energy landscapes.~\cite{Laio2008,Bernardi2015,Valsson2016,Camilloni2018,Henin2022a}
Two of the most widely used families of enhanced sampling algorithms are based on the definition of a set of \textit{collective variables} (CVs) that approximate the reaction coordinate~\cite{Laio2002} or on \textit{paths} that connect two end states of the process of interest.~\cite{Bolhuis2002} 
Each of these families of methods have their strengths and limitations. 

CV-based algorithms crucially depend on the identification of slow degrees of freedom that capture the transition under investigation and are typically limited to biasing a maximum of three variables at the same time.~\cite{Pietrucci2017,Palacio-Rodriguez2022}
Path-based methods can collect a rather large number of degrees of freedom into a feature space to be used in the path-building process. However, the effectiveness of this process generally depends on a proper definition of the end states and the effectiveness of the path search algorithm.
The original \textit{Path-collective variables} (PATHCVs) combine in an effective manner various aspects of the two families of approaches.~\cite{Branduardi2007} They are typically used in combination with an enhanced sampling algorithm and thus they can be used to explore paths that are separated by high free energy barrier from the initial (guess) path. Because of their efficiency in exploring path space, PATHCVs have been successfully used to reconstruct the free energy profiles associated with many complex systems.~\cite{Berteotti2009,Fidelak2010,Fribourg2011,Saladino2012,Cignoni2021} In addition, their combination with partial path transition interface sampling was shown to produce accurate numerical prediction of rate constants.~\cite{Juraszek2013}
Recent developments include approaches where the path is updated adaptively~\cite{Ortiz2018} or where multiple paths are taken into account simultaneously.~\cite{Ortiz2021}

Still, the effectiveness of PATHCVs depends on the choice of the metric used to define the path, as well as on milestones number, position and other parameters. We have recently explored a machine learning approach based on a spectral gap optimization procedure to find optimal linear combinations of different features for the metric of PATHCVs.~\cite{Hovan2019}
However, the approach is limited by the fact that a fixed linear combination of different features not always provides an optimal metric for all the milestones defining a complex paths. A typical example of this is provided for instance by the binding of a ligand to a protein. In such a case, the role of ligand hydration and dehydration is crucial at a given milestone (when the ligand enters or exits the cavity), while it is less relevant in the remaining path (corresponding to the solvated ligand being free in solution or finding its way to the cavity). This was clearly shown by quantifying the relevance of various descriptors, including ligand hydration, at different stages of ligand binding.~\cite{Rizzi2021} In that context, it was shown that a CV that combines in a non-linear way a number of descriptors~\cite{Bonati2020} is able to capture the relevant slow degrees of freedom (including hydration) at the right time. Thus, using a fixed metric to define the distance from the milestones defining the path can result in a sub-optimal performance.

In recent years, several methods have been proposed to learn CVs directly from data, using either supervised or unsupervised methods.~\cite{Ma2005,DiazLeines2012,Perez-Hernandez2013,Tiwary2016,Sultan2017,ribeiro2018reweighted,Mendels2018,Chen2018,Gkeka2020,Bonati2020,Bonati2021,Hooft2021,Trizio2021a,Sun2022,Ray2023} 
Some of the learned variables were shown to be close to paths connecting metastable states~\cite{Bonati2023,lelievre2023analyzing}, without this condition being explicitly imposed. Also, a data-driven variant of PATHCV based on kernel ridge regression has been proposed, in which the committor probability is approximated.~\cite{Pietrucci2023}

In this paper, we use machine learning (ML) techniques to overhaul the path CV approach and overcome the limitations associated with previous implementations. The resulting variable, called Deep-locally-non-linear-embedding (DeepLNE), offers a path-like-behavior when trained on a short timeseries of a system moving from A to B in phase space. It offers an automatic procedure of building an optimal 1D description of the training data, inspired by the formalism of locally linear embedding~\cite{Roweis2000} (LLE) and the PATHCV. Essential elements of the algorithm are a dimensionality reduction operated by an artificial neural network (ANN), a differentiable k-nearest-neighbor selection (k-NN) and an encoding of the neighborhood into a 1D latent space via an ANN.
Starting from a predefined feature set describing the system dynamics of interest, the DeepLNE algorithm creates a representation of each datapoint based exclusively on its neighbors in the training dataset. This metric is then transformed into a path-like CV locally anchored by the data provided during the training phase.

We test the DeepLNE CV on simulations of a particle in the M\"uller-Brown potential, alanine dipeptide, and an RNA 8-mer using different variants of the On-the-fly Probability Enhanced Sampling \cite{Invernizzi2020,Invernizzi2020d,Invernizzi2022,Rizzi2023} (OPES) method. We show that the application of DeepLNE is successful not only in the simplest examples but also in the cases of high-dimensional inputs or the combination of a set of heterogeneous features into an optimal path-like CV. 

\section{Methods}

Prior to introducing the DeepLNE method, we provide a brief overview of the PATHCVs and LLE frameworks, highlighting the areas of inspiration for our work.

\subsection{The path collective variables}

In the original PATHCVs approach~\cite{Branduardi2007}, one encodes a path connecting two end state A and B into a progress along the path CV $s$ and a distance from the path $z$. Given a $D$-dimensional feature space $\bm{X}$, one can express it parametrically $\bm{X}(t)$ along the path so that $\bm{X}(0) = \bm{X}_A$ and $\bm{X}(1) = \bm{X}_B$. Formally, the PATHCVs $s$  and $z$ are two functions of $\bm{X}(t)$ that are defined as
\begin{align}
\label{eq:s}
s(\bm{X}) &= \lim_{\lambda\rightarrow\infty} \frac{ \int^1_0 t \, \mathrm{e}^{-\lambda (\bm{X}-\bm{X}(t))^2} \mathrm{d}t} {\int^1_0 \mathrm{e}^{-\lambda (\bm{X}-\bm{X}(t))^2} \mathrm{d}t } \\
\label{eq:z}
z(\bm{X}) &= \lim_{\lambda\rightarrow\infty} - \frac{1}{\lambda} \ln \int^1_0 \mathrm{e}^{-\lambda (\bm{X}-\bm{X}(t))^2} \mathrm{d}t.
\end{align}

In the practice, this PATHCV formulation cannot be employed directly to deposit bias during simulations because of the crucial requirement of a CV to be smooth and differentiable. One has to resort to a number of assumptions and approximations. First, the parametric path is replaced by a discretised version $\bm{X}_i$ constructed on a series of $m$ path milestones, such that the integrals transform into finite sums. Choosing a number of high-quality milestones that are equidistant in $\bm{X}$ is a non-trivial task that requires empirical optimization.

Another challenge is the choice of $\lambda$. A $\lambda\rightarrow\infty$ leads to a step-wise path that is dominated by contributions for which $(\bm{X}-\bm{X}_i)^2$ is minimal and whose derivative is ill-behaved. A finite optimal $\lambda$ strikes a balance between high values corresponding to sharp PATHCVs and low values that sacrifice resolution in favour of differentiability.

An often overlooked challenge is the choice of $\bm{X}$ itself. For a PATHCV to be effective in enhanced sampling simulations, the feature space must contain as many as possible degrees of freedom so that the relevant ones pertaining to the transition between states A and B are captured. At the same time, the number of dimensions of $\bm{X}$ should be reasonably small, as the Euclidean norm used in Eq.~\ref{eq:s}-\ref{eq:z} quickly loses resolution for an increasing dimensionality of $\bm{X}$. Here the conundrum lies in navigating these two conflicting requirements which represent a complex optimization problem. Attempting to solve it requires finding a difficult compromise that is often problem dependent and cannot be easily generalized.

In order to construct the PATHCVs one typically follows a multi-step procedure. The starting point is a trajectory of the system of interest in which at least one transition from A to B is observed. Next, a feature vector $\bm{X}(t)$ is determined, that should discriminate as uniquely as possible the state of the system at any given time $t$. Then, one selects snapshots along the transition that define the path $\bm{X}$ and picks a finite $\lambda$. For reasons of computational efficiency, the number of snapshots $m$ must be much smaller than the number of sampled datapoints t. The milestone selection can be performed in a number of ways but is typically empirically crafted, requiring expert knowledge of the system and CV-building experience. We will see how some of these steps can be automated with the help of machine learning.

\subsection{The Locally Linear Embedding method}
\label{MM_LLE}

Locally Linear Embedding (LLE)\cite{Roweis2000} is a nonlinear dimensionality reduction technique that can be used to transform high-dimensional data into a lower-dimensional space with the constraint of preserving the local structure (i.e. the neighborhood) of the data. This is achieved in two steps.

Assuming that each data point $\bm{X}_{i}$ possesses $k$ neighbors that lie on a locally linear patch of its manifold, the LLE method reconstructs $\bm{X}_{i}$ from its neighbors via a linear regression. At first, one minimizes the cost function
\begin{equation}
\label{eq:e}
\epsilon(W) = \sum_i| \bm{X}_i - \sum_j W_{ij}\, \bm{X}_j |^2~,
\end{equation}
with the constraints that $\sum_j W_{ij} = 1$ and that each data point $\bm{X}_i$ is reconstructed exclusively from its neighbors. By doing so, one determines the optimal weight matrix $W$, where $W_{ij}$ represents the contribution of the $j$\textsuperscript{th} data point to the $i$\textsuperscript{th} reconstruction. For our purposes, this corresponds to assuming that each point in a trajectory used to generate a PATHCV can be represented locally by $k$ other points.

In a second step, LLE builds a lower-dimensional neighborhood-preserving mapping in such a way that the local linear relationships are approximately preserved. This implies that the same weights $W_{ij}$ that reconstruct the $i$\textsuperscript{th} data point in $D$ dimensions should be able to reconstruct also its embedded manifold coordinates $\bm{x}_i$ (of dimension $d$). This is accomplished by minimizing the embedding cost function:
\begin{equation}
\label{eq:phi}
\phi(\bm{x}_i) = \sum_i|\bm{x}_i - \sum_j W_{ij} \,\bm{x}_j|^2 ~,
\end{equation}
where $W_{ij}$ is now fixed.

Constructing a PATHCV by applying the standard LLE method still involves the application of the Euclidean metric on the feature space (compare Eq.~\ref{eq:s} and Eq.~\ref{eq:phi}). The higher dimensional the space, the more the computed Euclidean distances and the low-dimensional mapping suffer from degeneracy, i.e. a number of different feature space vectors produce analogous compressed values. Therefore in both contexts, one faces the same challenge: the choice of an optimal metric. 
Furthermore, while for the PATHCVs one has to define the total number of $m$ milestones, in LLE one has also to make an \textit{a priori} choice of the number of neighbors $k$ that locally represent each datapoint.

\subsection{The Deep-locally-non-linear-embedding method}
\label{DeepLNE_method}

\begin{figure*}[htb]
\includegraphics[width=\textwidth]{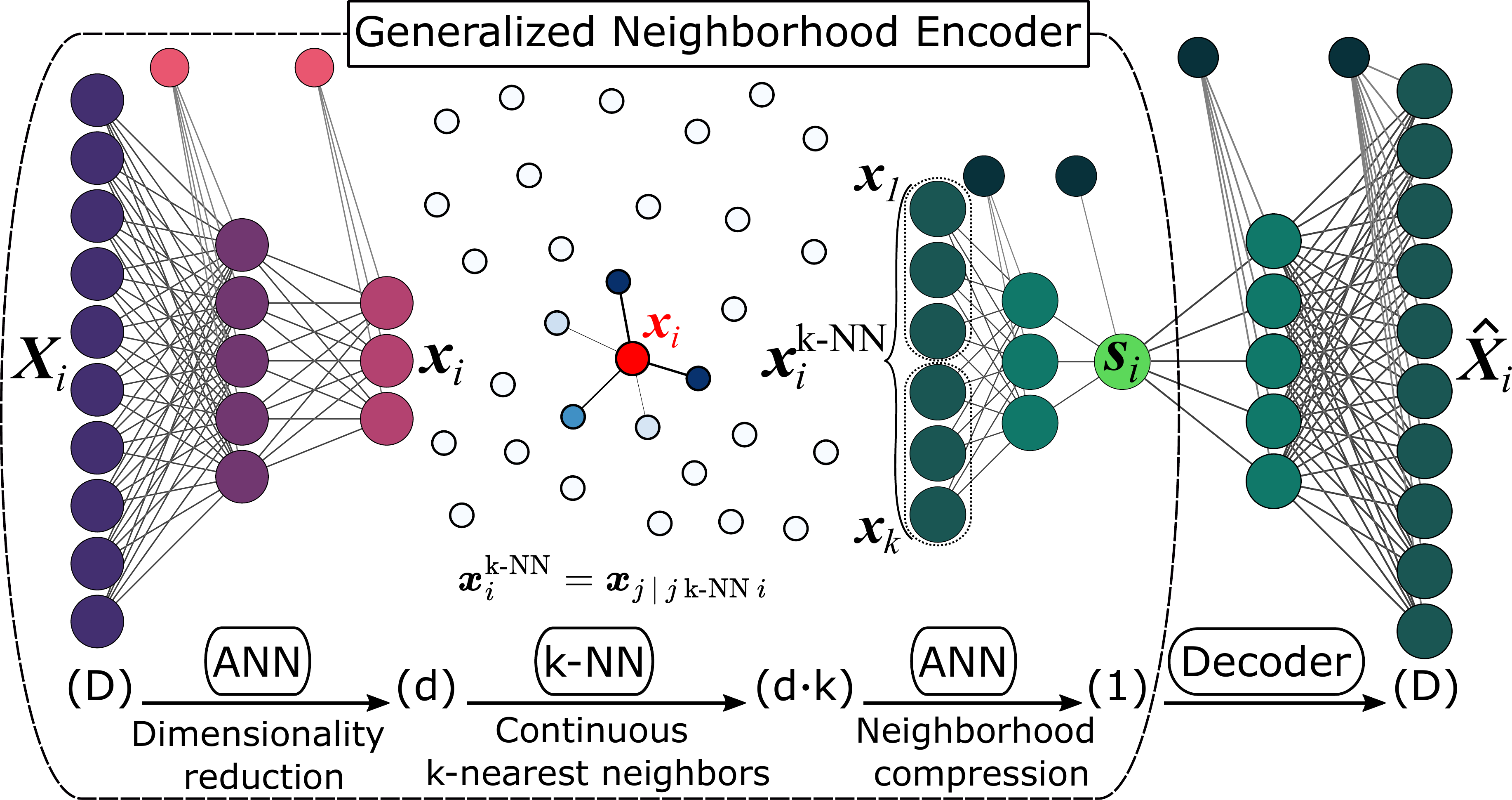}
\caption{\label{fig:SchematicCV} Schematic of the DeepLNE CV architecture. We construct a 'Generalized Neighborhood Encoder' by using an ANN, performing a first dimensionality reduction, a differentiable $k$-nearest-neighbor selection ('k-NN') followed by another ANN in order to compress the identified nieghborhood representation into a single dimension, the $s$ variable. The original choice of input features $\bm{X}_i$ for a single datapoint (out of $m$ total training data) with dimensionality D is reduced to a $d$-dimensional vector $\bm{x}$.  In this non-linearly transformed manifold, for each datapoint, $k$-nearest neighbor datapoints are selected with a weight matrix based on the Euclidean distance. 
We combine the features $\bm{x}_j$ of $k$-nearest-neighbors to construct the vector $\bm{x}_i^\mathrm{k-NN}$. This vector represents the input to a subsequent ANN that is used to compress the neighborhood into a 1D vector, the DeepLNE variable $s$, denoting the progression along the path. The decoder tries to optimally reconstruct the input $\bm{X}$ from $s$, resulting in vector $\hat{\bm{X}}$. The perpendicular distance from the path $z$ is then computed as a function of the Euclidean distance from $\hat{\bm{X}}$. The flowchart below the network architecture reports the vector dimensionality of the single training datapoint at each step along the DeepLNE CV construction.
}
\end{figure*}

The DeepLNE CV that we develop here (see Fig.~\ref{fig:SchematicCV}) aims at representing a high dimensional dataset with a single dimension, by incorporating concepts from PATHCV and LLE into a neural network architecture.
Its starting point is a trajectory of a system going from state A to state B and a set of physical features $\bm{X}$ evaluated along it. Its objective is to build a directional CV $s$ that can describe the progress along the transition through a non-linear combination of such a feature vector.
The architecture of DeepLNE can be seen as a generalized autoencoder, with the first part being composed by an artificial neural network (ANN) to perform a preliminary dimensionality reduction, followed by a continuous k-nearest neighbor (k-NN) step on each datapoint, which is then passed to an encoder that compresses the neighborhood into a one-dimensional representation. This 1D CV $s$ is then used to reconstruct the original input via a decoder, that is also used for computing an accessory perpendicular distance $z$ CV. 

This strategy is built upon the satisfaction of a number of constraints. The first one requires that the initial $m$ datapoints $\bm{X}$ can be optimally reconstructed from their one-dimensional representation $s$. We set this restriction in place with an asymmetric autoencoder architecture where the reconstruction loss between the original data $\bm{X}$ and the decoded data $\hat{\bm{X}}$ is minimized.

Second, in analogy with the PATHCVs and LLE, each datapoint $i$ must be exclusively represented by its neighbors. In order to satisfy this constraint, we implement a continuous and differentiable relaxation of the k-NN selection rule (see Sec.~SI~1 and Ref.~\onlinecite{Tobias2018}). Notably, the compressed latent space representation $s_i$ is not directly obtained from a transformation of the input features of a datapoint $\bm{X}_i$, but from those of its $k$ neighbors. The resulting CV is thus more robust to extrapolation as it is anchored to the local description of the training data points at all times.   

As a third constraint, the neighbors of each datapoint are found in a different manifold of arbitrary dimension, in our case in a lower dimension $d$ compared to the original feature space dimension $D$. This is achieved by transforming the initial inputs using an ANN. The PATHCV and LLE methods discriminate datapoints by applying the Euclidean distance directly on the original feature space, therefore their compressed representation may suffer from a degeneracy issue for large $D$. Instead DeepLNE computes Euclidean distances exclusively in the reduced dimensional space $d$ during the k-NN step, thus alleviating the degeneracy issue.

The fourth and last constraint of DeepLNE is to represent each datapoint via a non-linear combination of the neighbors' features. 
This aspect is also a novelty, as PATHCV and LLE exclusively rely on linear transformations of the original datapoints.
We expect this increased flexibility to play a major role in cases where the sampled data are sparse, such as in high dimensions or in the vicinity of sampling bottlenecks.

In analogy with the seminal PATHCV framework, the DeepLNE CV can provide a measure of both the progress along the path ($s$) and the distance from it ($z$), which can be used for driving the system along the process and controlling it. While $s$ is computed as the output of the encoder part of the DeepLNE architecture described above, the $z$ variable is defined from the decoded $s$ variable. For a new datapoint $\bm{Y}$, the perpendicular distance to the path $z$ is computed as 
\begin{equation}
\label{eq:zDeepLNE}
z(\bm{Y}) = \lim_{\lambda\rightarrow\infty} - \frac{1}{\lambda} \ln \left( \sum_i^{m}{e}^{-\lambda (\bm{Y}-\bm{\hat{X}_i})^2} \right)~,
\end{equation}
which is analogous to Eq.~\ref{eq:z}, where $\bm{X(t)}$ is replaced by the decoded training data $\bm{\hat{X}}$ that are schematically depicted in Fig.~\ref{fig:SchematicCV}. By comparing the new data not directly with the training datapoints, but with their decoded versions (i.e., passed through neighbor selection and then reconstructed), we can more effectively measure the distance from the average learned path. Note that, in this definition, the $z$ variable might still suffer from the high dimensionality problem of $\hat{\bm{X}}$. In this regard, we checked whether the variable $z$ can be calculated with respect to the reduced dimension $\bm{x}$ instead of $\bm{\hat{X}}$. However, we found in the numerical examples that the $z$ computed in this alternative way did not improve upon our recommended definition Eq.~\ref{eq:zDeepLNE}, and in some cases, it even decreased the sampling efficiency along the transition of interest.

In summary, the DeepLNE algorithm makes it possible to construct a variable that approximates a transition path between states, while overcoming multiple issues of PATHCVs:
\begin{itemize}
\item A lower dimensional metric $d$ is learned automatically and reduces the degeneracy when computing the Euclidean norm to identify the local neighborhood of a datapoint.
\item Local descriptions of the high dimensional dataset do not have to be empirically chosen, but are found in an automatic way by constructing the neighborhood of each data point using a differentiable k-NN step. 
\item Heterogeneous input features are permitted, \textit{e.g.} we can combine distances, angles and contact maps into $\bm{X}$. 
\end{itemize}

\subsection{Differentiable nearest neighbors selection}
An important ingredient for the deployment of the DeepLNE CV in the enhanced sampling context is the adoption of a continuous and differentiable approximation of the k-NN selection rule. While the technical details of the implementation are given in Section ~SI~1, here we focus on the choice and the interpretation of two hyperparameters: the number of neighbors $k$ and the sparsity of the selection matrix $t$. 
Concerning the number of neighbors $k$, we have the following limiting cases:

\begin{itemize}
\item$k\rightarrow 1$: leads to sharp local description of training data and high variability of the DeepLNE CV when extrapolating out-of-distribution. This is due to the fact that each datapoint is exclusively described by a single neighbor.
\item$k\rightarrow\infty$: in this limit, we have a loss of locality and therefore loss of path-like behavior, especially in the regions that are sparsely described or under-sampled, such as the transition state region which is crucial for capturing the dynamics of the process. Furthermore, the inter- and extrapolation of variables $s$ and $z$ become more degenerate, as a given datapoint is described by many training samples. 
\end{itemize}
The second hyperparameter $t$ determines the smoothness of the k-NN step by adjusting whether only the first $k$ neighbors or a larger set of neighbors contribute (see Fig.~SI~1). Its role is comparable to the one of $\lambda$ in the PATHCV framework. We single out the following limiting cases:
\begin{itemize}
\item$t\rightarrow0$ imposes exact nearest neighbor selection in which a datapoint is exclusively described by the features of the selected neighbors. This behavior is problematic in biased trajectories where, during the phase space exploration in $s$, the identity of the neighbors would sharply change causing discontinuities in its derivative. The effect is analogous to the case where $\lambda\rightarrow\infty$ for the PATHCV. 
\item$t\rightarrow\infty$ implies that a datapoint is described by a function of all the training datapoints, strongly lowering the resolving power of $s$, in analogy with $\lambda\rightarrow 0$ for the PATHCV. 
As a consequence, each new point gets transformed into a point in the center of the training set.
\end{itemize}
Based on these limiting cases, we suggest that $k$ should be rather small ($k < 10$), to retain the path-like behavior of the CV, while $t$ should be small but finite ($t \approx 0.1$) to ensure numerical stability during enhanced sampling simulations.

\subsection{Training the DeepLNE CV}

For all the systems studied in this paper, we followed the same protocol: 1) sampling of the transition of interest, 2) training the DeepLNE CV on a set of input features, 3) performing enhanced sampling MD simulation using the DeepLNE CV for bias deposition. 
In the following, we outline all the key steps of our recommended strategy for training the DeepLNE CV.

\textsf{\textbf{\small Training data generation.}} Starting trajectories must describe fully and as dense as possible the transition from state A to B. They can be unbiased in the simplest of cases where the energetic barrier between states is small or they can be biased data. Among the biasing procedures to generate input trajectories, we would single out the ratchet-and-pawl restraint \cite{Camilloni2011,ABeccara2012}, steered MD \cite{Grubmuller1996} and OPES in exploration mode \cite{Invernizzi2022}. Since the computational cost of DeepLNE increases with an increasing number of datapoints, one can rely on Farthest Point-Sampling (FPS)\cite{Goscinski2023} to select a maximally diverse subset of the original data, such that only $m$ frames are retained. 

\textsf{\textbf{\small Neural network architecture}}. In the examples reported in this manuscript, we use a 2 feed-forward ANN both containing at most 2 hidden layers and the hyperbolic tangent activation function. The ANN architecture can be adjusted as desired by the user. We recommend choosing $d$ such that dimensional reduction per hidden layer and the total number of fitting parameters strike a balance. While the expected application scenario for complex biological systems of many degrees of freedom is $d << D$, cases in which $d > D$ can also be envisioned. Furthermore, in order for the $s$ variable to describe the progress along the path, we scale it using a sigmoid activation in the final encoder layer, restricting its values to the range of 0 to 1. 

\textsf{\textbf{\small Neural network optimization}}. The network parameters are optimized using the mean square error between the original input and the reconstructed ones:
\begin{equation}
    \mathcal{L}=\frac{1}{m}\sum_{i=1} ^m | \bm{X}-\bm{\hat{X}} |^2
\end{equation}
via gradient descent using the ADAM optimizer with a learning rate of $10^{-3}$ and 5000 epochs, using the machine learning library PyTorch \cite{Paszke2019}, with the help of the \verb|mlcolvar| package\cite{Bonati2023}. 

\textsf{\textbf{\small Inspecting the results}}.  Furthermore, it can be instructive to plot the DeepLNE CV along a few important physical descriptors or against the first principal components of the input features $\bm{X}$. A more detailed analysis can be obtained via a sensitivity analysis to identify the features that contribute the most{\cite{Bonati2020}} or by using linear models to interpret the neural network-based CV.~\cite{novelli2022characterizing}

\textsf{\textbf{\small Exporting the model into PLUMED}}.
After the training is finalized, the model is compiled using TorchScript. The DeepLNE CV is evaluated on the fly in MD simulations by loading the model into the PLUMED plugin~\cite{Tribello2014} as a PYTORCH CV.~\cite{Bonati2023} An implementation of the method and usage tutorials are available at \url{https://github.com/ThorbenF/DeepLNE}.

\subsection{Simulation details}
\label{SimulationMM}
\begin{figure*}
\includegraphics[width=\textwidth]{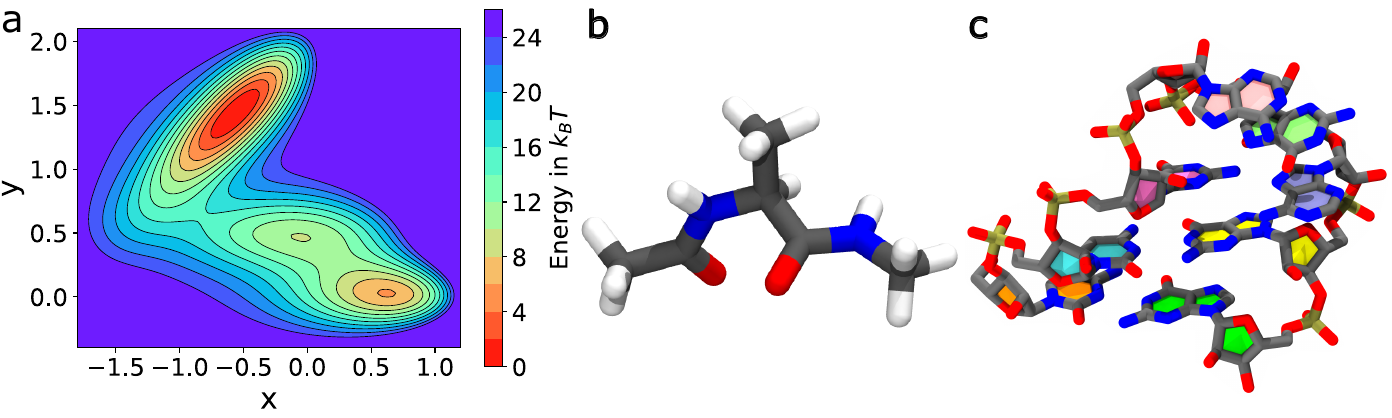}
\caption{\label{fig:SchematicSystems}
The DeepLNE CVs are tested on 3 toy models: (a) M\"uller-Brown potential used as a 2-state potential energy landscape for the simulation of a particle moving in two dimensions. (b) Structure of the well-studied biomolecule alanine dipeptide, a system that can be sufficiently described via its dihedral angles $\phi$ and $\psi$. (c) Structure of an RNA tetraloop made up of 8 nucleotides (the RNA with sequence 'CCGAGAGG' is coloured in 5' to 3' direction: orange, cyan, purple, pink, lime, yellow, green). We show the correctly folded structure.}
\end{figure*}

To test the introduced algorithm, we report simulation results for a particle in the M\"uller-Brown potential, alanine dipeptide and an RNA tetraloop (Fig.~\ref{fig:SchematicSystems}). 

\textsf{\textbf{\small M\"uller-Brown.}} The simulations of a particle subjected to the M\"uller-Brown potential are performed using the simple Langevin dynamics code contained in the VES module of PLUMED~\cite{Valsson2014}, and the biased simulations are performed using the OPES~\cite{Invernizzi2020} method with a pace of 200 steps, the automatic bandwidth selection, and a barrier parameter equal to 10 $k_BT$. The particle is moving in two dimensions under the action of the three-state potential depicted in Fig.~\ref{fig:SchematicSystems} built out of a sum of Gaussians. During step 1) we choose to employ the ratchet and pawl ABMD method~\cite{Camilloni2011} biasing the y-coordinate to sample the transition between 2 states A and B. Then in step 2) the DeepLNE CVs ($s$ and $z$) are trained using the $x$ and $y$ position of the particle as input features. For step 3) a new simulation is run where $s$ is used as a CV to deposit bias via OPES. Also a harmonic constrained with a cutoff value of 0.02 (UWALL) is applied on the $z$ variable. We estimate statistical errors via block analysis using 3 blocks.

\textsf{\textbf{\small Alanine dipeptide}}. For the extensively studied FES of alanine dipeptide, 3 transition paths are investigated considering the dihedral angles $\phi$ and $\psi$. During step 1) we apply ABMD biasing both dihedrals so that 5 independent transitions across the FES barrier of interest are observed. 
To test the dimensionality reduction capability of DeepLNE, we also employ all inter-atomic distances in the molecule as initial features, which corresponds to $D=190$.  
To save computational costs we use the FPS tool of Ref.~\onlinecite{Goscinski2023} to reduce the number of training datapoints ($m=3000$). After training the DeepLNE CV, we use the $s$ variable as a CV to bias in combination with OneOPES~\cite{Rizzi2023} and 8 replicas. We choose a pace of 500 steps, the automatic bandwidth selection, and a barrier parameter equal to 60 $k_BT$ and 2.5 $k_BT$ respectively for OPES-METAD and OPES-EXPLORE. Also a harmonic constrained with cutoff value and $\kappa = 20000$ (UWALL) is applied on the $z$ variable. Simulations are carried out with the DES-Amber ff.~\cite{Piana2020} We analyze replica $0$ and estimate statistical errors using blocks of 2 ns.

\textsf{\textbf{\small RNA tetraloop}}. We select an RNA 8-mer as a model for a system with biological relevance and configurational complexity. The system consists of 8 nucleotides with the sequence 'CCGAGAGG' containing the GAGA Tetraloop motif (Fig.~\ref{fig:SchematicSystems}) and has been previously studied using extensive enhanced sampling MD \cite{Bottaro2016}. We use the FES obtained by using 24 replicas with a simulation length of 1 $\mu s$ per replica for comparison. As proposed by the authors in this previous study we will include recent corrections to the van der Waals parameters and a more accurate water model.
Consequently for our simulation we used the standard OL3 RNA ff~\cite{Cornell1996,Wang2000,Perez2007,Zgarbova2011}  with the van-der-Waals modification of phosphate oxygens developed in Ref.~\onlinecite{Steinbrecher2012} without adjustment of dihedral parameters. As a water model we chose OPC.~\cite{IzadiOPC2014}
This combination has been originally proposed in Ref.~\onlinecite{bergonzo2015improved} and  already tested on RNA systems.~\cite{bergonzo2022conformational,Froehlking2023}
The simulations were run at salt concentrations corresponding to a system with neutralised total charge using the Joung-Cheatham ion parameters~\cite{CheathamIons2008} optimized for TIP4PEwald. To sample the transition from natively folded state A to unfolded state B we use the ABMD method biasing 3 different CVs: 1) the eRMSD metric \cite{Bottaro2014} with respect to state A, 2) a contact map constructed using the heavy atom distances between the 4 nucleotides in the stem 1,2 and 7,8 with cutoff 5 $\angstrom$ (CMAP) inspired by Ref.~\onlinecite{Rahimi2023} , 3) a contact map constructed using the heavy atom distances between the 4 nucleotides in the loop 3,4 and 5,6 with cutoff 3.5 $\angstrom$ (CMAP). We start ABMD runs from the unfolded state at an eRMSD $>2.1$ and stop the simulation upon reaching the natively folded state at an eRMSD $<0.2$.
We reduce the number of training datapoints ($m=2250$) via FPS and choose the same 3 CVs used for the ABMD simulations as input features in the DeepLNE training. In the enhanced sampling simulation using DeepLNE we use the $s$ variable to deposit bias with OneOPES. For OPES-EXPLORE we choose a PACE of 20000 steps, automatic bandwidth selection, and a barrier parameter equal to 50 $k_BT$. The temperatures for the 8 replicas are selected from a distribution ranging from 300 K to 400 K and each replica is simulated for 500 ns.
As done in the comparison study of the GAGA-Tetraloop\cite{Bottaro2016}, replica $0$ is analyzed excluding the first 200 ns. Statistical errors were estimated using blocks of 100 ns.

Simulations of Alanine dipeptide and GAGA-tetraloop are performed using the GROMACS 2022.5 engine~\cite{abraham2015gromacs} patched with the PLUMED 2.9 plugin~\cite{Tribello2014} with the Hamiltonian replica exchange algorithm.~\cite{Bussi2014} 

After simulations of the toy models M\"uller-Brown and the alanine dipeptide were completed, we performed a committor analysis to investigate whether the identified DeepLNE variable $s$ approximates well the real reaction coordinate. For the M\"uller-Brown potential the transition region is split into 5 bins along the $s$ CV in the region of the maximum in the FES and for each bin 300 particle positions are extracted and independent simulations are performed. In the case of alanine dipeptide, we split the FES of replica $0$ in the vicinity of the transition state region into 10 bins along the $s$ variable and extract 100 structures with the lowest values of the $z$ variable. We then start new independent simulations from these structures, bookkeeping which state they visit first (A or B).

\section{\label{sec:Results}Results}

\subsection{Particle in M\"uller-Brown-Potential}
\label{Muller_Brown_Potential}

\begin{figure*}[htb]
\includegraphics[width=\textwidth]{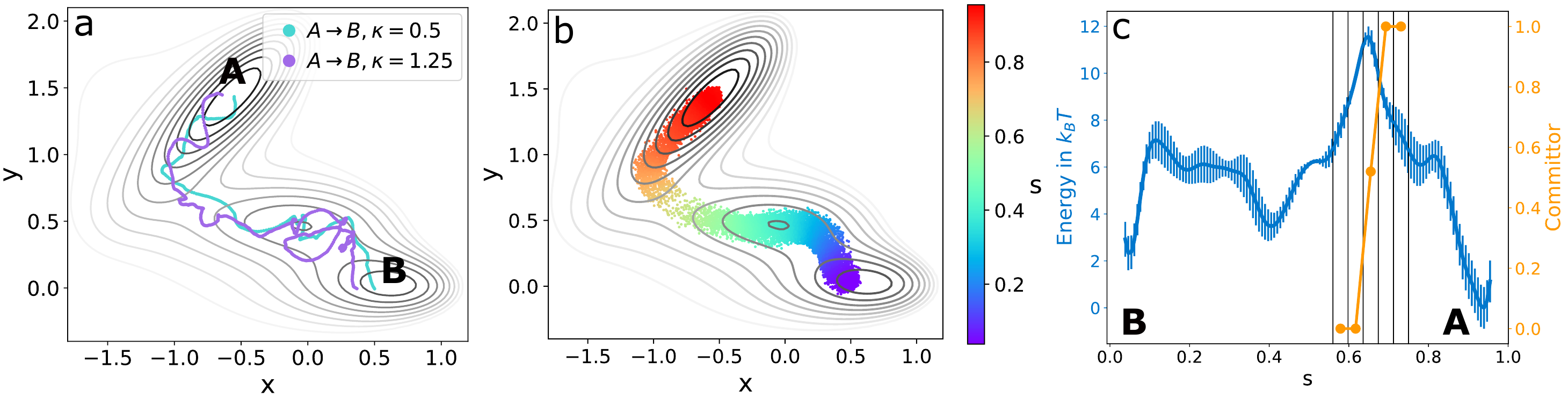}
\caption{\label{fig:MullerBrown} Results of training DeepLNE for a particle simulation in the M\"uller-Brown-potential.
(a) Training datapoints for DeepLNE, obtained using ABMD (biasing y-coordinate with different spring constants $\kappa$) describing the movement from state A to state B. 
(b) Sampled configuration during the biased simulation using OPES applied on the trained DeepLNE CVs $s$ and a harmonic constrained applied on z. The color of the datapoints correspond to the DeepLNE $s$ variable showing path-like behavior. 
(c) Free energy with respect to the $s$ variable as well as the committor probability for state A. The maximum of the FES and the committor value of $0.52$ coincide in the same bin for s.}
\end{figure*}

We demonstrate how to train and apply the DeepLNE CV to enhanced sampling MD simulations. In Fig.~\ref{fig:MullerBrown} we collected the results of the 3 steps along the proposed DeepLNE algorithm. 
We start by performing 2 simulations biasing the $y$-coordinate via ABMD using different spring constants $\kappa=0.5$ and $\kappa=1.25$. As can be seen from Fig.~\ref{fig:MullerBrown} (a), both runs start from the same state A and are stopped as soon as they reach state B. We show the portion of trajectory that describes the transition from state A to state B. We use this reduced set of datapoints to compute the input features for the DeepLNE CV training step. In Fig.~\ref{fig:MullerBrown} (b) we see the results of biasing the particle simulation using the $s$ variable of the DeepLNE model and constraining perpendicular movement via the $z$ variable. One can appreciate that the desired exploration moving from state A to state B is achieved and that $s$ describes the intended progression along the path. 

In Fig.~SI~2 (a), we show the decoded training datapoints $\hat{\bm{X}}$ of DeepLNE. Since all datapoints used for training the DeepLNE CV describe a transition across the same barrier, $\hat{\bm{X}}$ approximates well the path along the minimum in the M\"uller-Brown FES. Consequently applying a harmonic constraint on $z$ during enhanced sampling MD allows for frequent transitions while sampling the movement between the state A and B as shown in Fig.~SI~2 (b). 
We can appreciate that the entire spectrum of the $s$ variable is explored. Since we exclusively sample the transition and not the entire free energy landscape (Fig.~\ref{fig:MullerBrown} (b)) we can confirm that the applied harmonic constraint on $z$ is effective, limiting the exploration perpendicular to the path-like variable $s$. In Fig.~\ref{fig:MullerBrown} (c) we show the results of a free energy estimation as well as the committor analysis based on the estimated FES. We compute the committor probability for state A in the 5 different bins along $s$ in the region of the highest free energy. Along increasing values of the $s$ variable, one can appreciate that the committor value changes from 0 to 1 and takes a value of 0.52 in proximity of the maximum barrier value in the FES, indicating that the DeepLNE CV approximates well the ideal reaction coordinate.

\subsection{Alanine dipeptide}
\label{Alanine_dipeptide}

\begin{figure*}[htb]
\includegraphics[width=\textwidth]{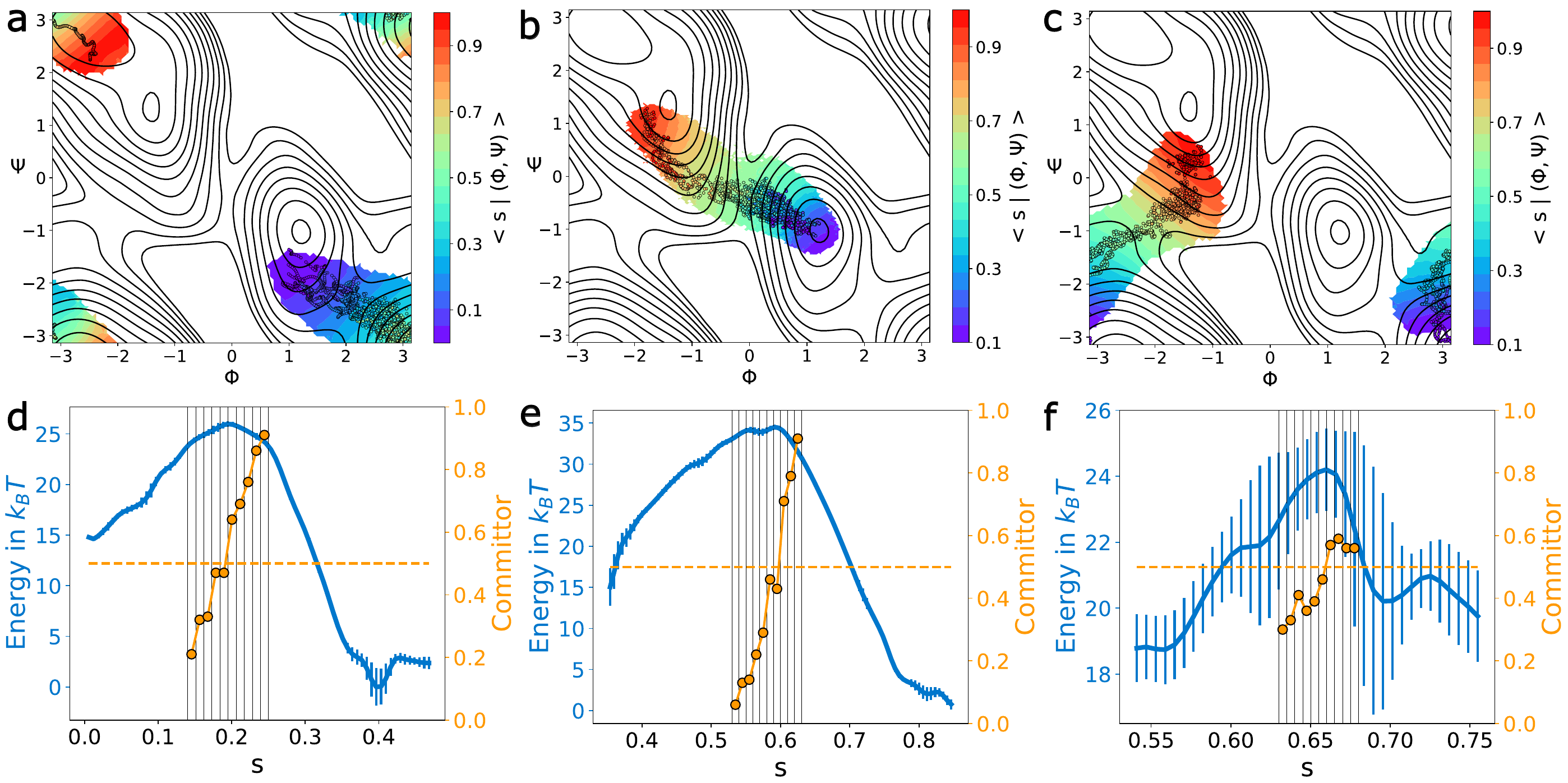}
\caption{\label{fig:Dialanine} Sampling of different pathways in alanine dipeptide using DeepLNE. In (a), (b) and (c), the scatter points represent ABMD trajectories crossing three distinct barriers. The DeepLNE variable is trained on those datapoint using 190 interatomic distances as input features instead of dihedral angles. Next, the DeepLNE CVs are employed to run OneOPES simulations and the new data is used to calculate the progress along the path, computed through the expectation value $<s | (\phi,\psi)>$ shown in the colored regions. The black lines below denote the reference FES, obtained by biasing $\phi$ and $\psi$ through OneOPES. The basin at $\phi=-1.5, \psi=2$ is referred to as $C_{7eq}$, while the basin at $\phi=1, \psi=-1$ is referred to as $C_{7ax}$.
The new OneOPES runs yield FES estimates of which subsections around the transition states are shown in panels (d), (e) and (f). The FES maxima coincide with committor probabilities close to 0.5 computed by running short simulations from the configuration in the same $s$ bins highlighted as vertical lines. 
}
\end{figure*}

Fig.~\ref{fig:Dialanine} reports the results of constructing the DeepLNE CVs for 3 different transitions of alanine dipeptide and estimating the FES via OneOPES simulations using the $s$ variable to deposit the bias potential. We note that the DeepLNE CVs are a non-linear function of the 190 inter-atomic-distances in alanine dipeptide.
Fig.~\ref{fig:Dialanine} (a), (b), (c) show the datapoints initially sampled from short ABMD runs across 3 different barriers in the reference FES obtained using OPES-METAD biasing $\phi$ and $\psi$ over 20 ns. We color the FPS reduced training datapoints based on the DeepLNE variable $s$. The variable describes the intended progression from one state A to another state B across the different barriers in all 3 cases.

Having constructed 3 path-like variables, we estimate the FES performing OneOPES simulations along them and also perform a committor analysis for each case (Fig.~\ref{fig:Dialanine} (d), (e), (f)). Before going into more detail, we note that in Fig.~SI~3 we collected the expectation value of the $z$ variable as well as the time evolution of the $s$ variable. The analysis of $<z | (\phi,\psi)>$ (see Fig.~SI~3 (a), (b), (c)) confirms that path-like CVs are learned, that exhibit low $z$ values along the transition of interest. Importantly, frequent transitions occur and the entire spectrum of the $s$ variables is explored during the sampling of the 3 different transition paths (Fig.~SI~3 (d), (e), (f)).
Fig.~\ref{fig:Dialanine} a) superimposes 5 ABMD simulations starting from the same configuration in the state $C_{7eq}$ and then progressing to state $C_{7ax}$ over a barrier at $\phi=2, \psi=-2.5$. From the reference FES it can be seen that there exist a single transition state region, which is quantified as shown in Fig.~\ref{fig:Dialanine} d) by performing a OneOPES simulation. The transition region is located between $s$ values 0.14 to 0.25 and accordingly configurations from 10 bins in this range are extracted to perform a committor analysis. We find that the $s$ values in range 0.17 to 0.2 of the DeepLNE CV correspond to the transition state with committor values close to 0.5.

In Fig.~\ref{fig:Dialanine} b) we show another transition between the states $C_{7eq}$ and state $C_{7ax}$, this time across a barrier located close to $\phi=0, \psi=0$. The superimposition of the training data obtained from 5 ABMD runs and coloring them with the learned $s$ variable shows that the transition should be expected for $s$ values in the range 0.53 to 0.63 and accordingly we analyze the corresponding OneOPES simulation in Fig.~\ref{fig:Dialanine} e). It turns out that in this case $s$ values in the range between 0.58 and 0.6 lead to committor probabilities close to 0.5, allowing us also here to identify the configurations corresponding to the transition state. Finally, in Fig.~\ref{fig:Dialanine} c) we train the DeepLNE model on a third possible transition path in the alanine dipeptide FES. We can see for this case that the system starts from configurations of the state $C_{7eq}$ with $\phi=-1.5, \psi=0.5$. The transition along a transition region close to $\phi=-2, \psi=-1$  is considered completed when $\phi=3, \psi=-3$ is approached. Performing also in this case OneOPES simulations using the DeepLNE $s$ variable as the biased CV leads to the FES shown in Fig.~\ref{fig:Dialanine} f). The maximum of the free energy occur for $s$ values close to 0.65 and through the committor analysis in the range 0.63 to 0.68 we find committor probabilities close to 0.5 for $s$ values 0.65-0.66.
These results show that despite the challenge of being trained on high-dimensional input features the DeepLNE algorithm can successfully guide a simulation from A to B in free energy space and help identify the configurations belonging to the transition state.

Additionally, the relevance of the input features $\bm{X}$ can be estimated by accumulating the gradients of $\bm{X}$ with respect to the DeepLNE $s$ variable. This is especially insightful in a scenario where the input feature dimension is very high. We report this analysis in Fig.~SI~4 showing that for all 3 path-like DeepLNE CVs there exist differences in relevance between input features (Fig.~SI~4 (a), (b), (c)). Importantly the most relevant inputs identified are sensitive to the relevant dihedral angle changes required to perform the respective transitions (Fig.~SI~4 (d), (e), (f)). 

\subsection{RNA GAGA-tetraloop}
\label{RNA_GAGA_tetraloop}

\begin{figure*}[htb]
\includegraphics[width=\textwidth]{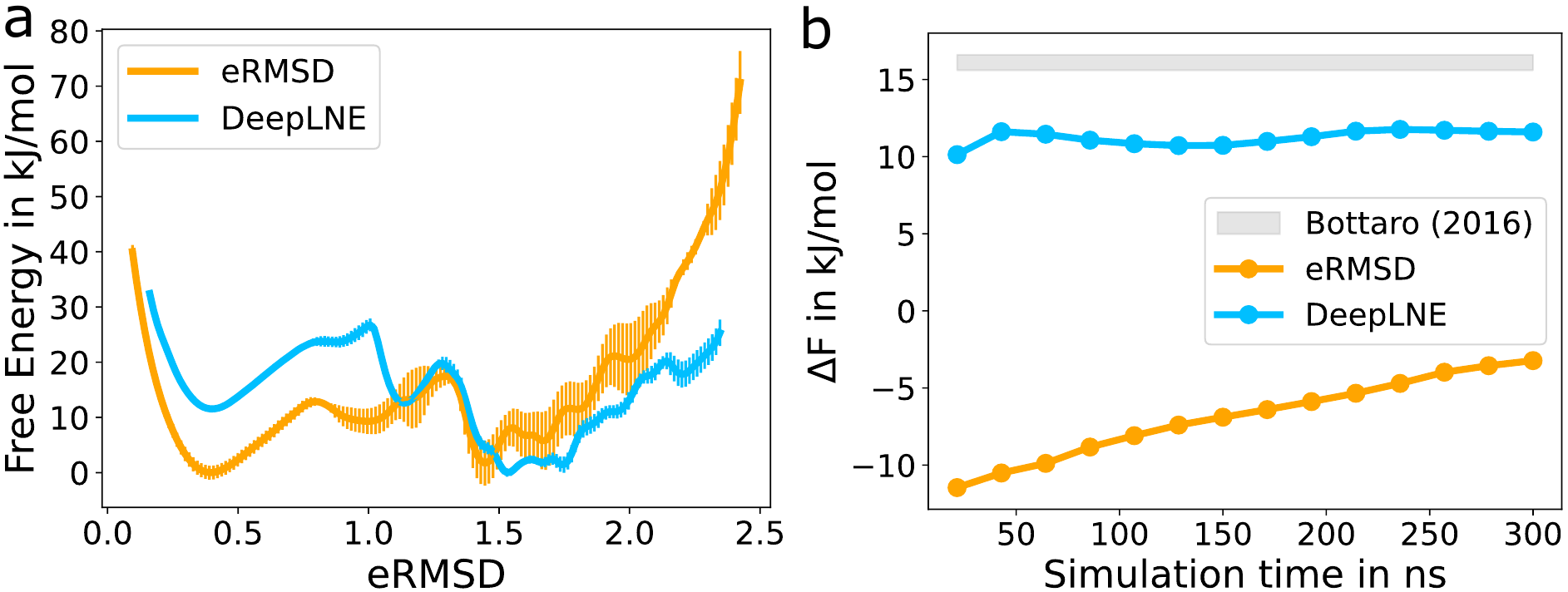}
\caption{\label{fig:GAGATetraloop} Free energy surface and free energy of folding ($\Delta F$) for the GAGA-Tetraloop based on 500 ns of OneOPES simulations using different biasing CVs (eRMSD, DeepLNE variable $s$). (a) Comparison of the FES along the eRMSD CV with respect to the natively folded conformation.
(b) Comparison of the $\Delta F$ over simulation time comparing the OneOPES simulations using 8 replicas to a previous study of Ref.~\onlinecite{Bottaro2016}, in which also the eRMSD was used as the biasing CV, but instead 24 replicas were simulated for 1 $\mu s$ each. Also a different force-field and TIP3P water was used instead.
}
\end{figure*}

In Fig.~\ref{fig:GAGATetraloop} a) we compare the FES estimation of the RNA GAGA-Tetraloop after 500 ns of simulation using OneOPES in combination with eRMSD or DeepLNE variable $s$ as the CV for bias deposition. As in the study by Ref.~\onlinecite{Bottaro2016}, the eRMSD is computed with respect to the natively folded conformation. The DeepLNE CV is trained on the eRMSD and 2 contact maps providing additional information about the contacts formed between the 4 nucleotides in the stem and the loop respectively. 

Our OneOPES simulation biasing the eRMSD for 500 ns results in a FES estimate that is in disagreements with previous studies where extensive enhanced sampling simulations were carried out. The FES in regions of conformations corresponding to high eRMSD values exhibit high statistical errors and are strongly underestimating the propensity of these conformations with respect to what has been previously observed (compare FES in Ref.~\onlinecite{Bottaro2016}).
Instead the OneOPES simulation using a trained DeepLNE variable $s$ to deposit a bias potential over time results in a FES which shows the expected relative population between natively folded and misfolded states. The statistical errors over the entire FES are consistently low.
The source of high statistical uncertainty in the case of the simulations biased via eRMSD can be seen in Fig.~SI~5. We compare the exploration of the biased CV (eRMSD, DeepLNE) over the course of the simulation. The analysis confirms that the eRMSD is not reliably exploring its entire range of values instead for simulations with the DeepLNE $s$ variable the entire spectrum of $s$ values is explored and frequent transitions occur. 

In Fig.~\ref{fig:GAGATetraloop} b) we report the estimated free energy of folding over the last 300 ns of simulation time.
In agreement with what has been discussed above the OneOPES simulation biasing the eRMSD is starting with a $\Delta F$ very different from previous studies and shows a drift over the course of the simulation. Instead, the OneOPES run using the DeepLNE CV $s$ for bias deposition already starts with an estimate of the folding free energy close to $16.1\pm 0.5$ kJ/mol found by Ref.~\onlinecite{Bottaro2016}. The $\Delta F$ estimate plateaus around $11.5$ kJ/mol. 
Since our simulations are performed using a force-field including van-der-Waals modification of phosphate oxygens and more accurate OPC water model, which were both not used in the study by Ref.~\onlinecite{Bottaro2016}, our $\Delta F$ estimation is expected to be similar, but not identical to the previous findings. We note specifically for the GAGA-tetraloop system, that even though experimentally the folded state is expected to be energetically favored in simulations the opposite is observed in simulations because of force field inaccuracies. Interestingly our simulation performed with the corrections envisioned by Ref.~\onlinecite{Bottaro2016} indeed lowers the free energy of folding. Nevertheless, the misfolded state still represents the minimum in the FES.

\section{\label{sec:Discussion}Discussion}

In this work, we propose a strategy to construct a path-like CV using a data-driven approach. The DeepLNE method builds a 1D representation of reactive trajectories that can be used to capture and especially accelerate rare events, with a focus on biophysical systems. Through multiple examples, we have shown how this CV approximates well the ideal reaction coordinate and can be effectively used to improve sampling using enhanced sampling methods such as OPES or its recently developed OneOPES variant.

Our methodology begins with obtaining a first reactive trajectory that captures the transition from state A to state B. Herein, these starting trajectories have been obtained with a ratchet-and-pawl restraint~\cite{Camilloni2011}, which provided training data of sufficient quality for all systems investigated. For more complex systems, an input reactive trajectory that samples the transition state regions more extensively might result in a DeepLNE CV that captures the region of the free energy barrier more precisely. To this end, an iterative biasing procedure such as the one introduced in Ref.~\citenum{Kang2024} might be helpful.

The input trajectory is used to define a feature vector to distinguish not only state A from state B but also the intermediate states encountered. With this featurized trajectory, one can automatically derive path-like variables applying the DeepLNE CV method introduced in this study. Input features are initially subjected to dimensionality reduction. In this lower dimensional manifold we compute Euclidean distances to search for a small number of k-nearest neighbors that describe the local neighborhood of each datapoint. Compressing this neighborhood representation even further finally yields the DeepLNE variable $s$, which, upon decoding, can be used to generate in turn the DeepLNE CV $z$. The interpretation of $s$ and $z$ is analogous to the PATHCVs, as they describe the progress along the path and its perpendicular distance, respectively. 

The proposed method alleviates the issue of CV degeneracy that occurs when calculating Euclidean distances in a high dimensional space. At the same time, it automates the process of constructing the CVs, since the $s$ and $z$ CVs are generated without making empirical choices such as selecting a set of milestones or a metric that have a strong impact on the quality of the CV and on the time required to fine-tune it. Unlike standard neural network-based CVs, the DeepLNE CV exclusively represents new data points through their neighbors in the training set. This feature is fundamental in preventing erroneous extrapolations when unknown data regions are visited.

Our method is not limited to two state systems and can be applied as is to more complex systems characterized by multiple metastable states. A DeepLNE CV can be successfully trained as long as the training trajectories visit all the states of interest and their transition regions. In the systems studied here, we did not impose a directionality on the CV in order to show that path-like CVs can be successfully derived even in a completely unsupervised manner. If needed, this can be ensured by adding a supervised term to the optimization (such as cross entropy, or a discriminative term), which can be done straightforwardly in a multi-task framework.~\cite{Bonati2023} 

All in all, we believe that DeepLNE provides a powerful and flexible method to automatically create efficient path-like CVs to accelerate sampling of complex phenomena.

\section*{Acknowledgements}
We acknowledge Simone Aureli for providing several useful suggestions. FLG, VR and TF acknowledge the Swiss National Supercomputing Centre (CSCS) for large supercomputer time allocations projectID:s1228. They also acknowledge the Swiss National Science Foundation and Bridge for financial support (projects numbers: $200021\_204795$, $CRSII5\_216587$ and $40B2-0\_203628$).
LB acknowledges funding from the Federal Ministry of Education and Research, Germany, under the TransHyDE research network AmmoRef (support code: 03HY203A). 

\section*{Data availability}
At \url{https://github.com/ThorbenF/DeepLNE} we provided the DeepLNE class with a corresponding tutorial together with scripts that are necessary to reproduce the figures and the results of this study.

\bibliography{main}

\newpage

\onecolumngrid

\clearpage

\begin{raggedright}
{\Large \textbf{Supplementary Material - Deep learning path-like collective variable for enhanced sampling molecular dynamics}\par}
\vspace{0.5cm}

\noindent
Thorben Fr\"ohlking$^{1,2,3}$, Luigi Bonati$^{4}$, Valerio Rizzi$^{1,2,3}$, and Francesco Luigi Gervasio$^{1,2,3,5}$\par
\vspace{0.3cm}

\begin{enumerate}
    \item \textit{School of Pharmaceutical Sciences, University of Geneva, Rue Michel Servet 1, 1206, Genève, Switzerland}
    \item \textit{Institute of Pharmaceutical Sciences of Western Switzerland (ISPSO), University of Geneva, 1206, Genève, Switzerland}
    \item \textit{Swiss Institute of Bioinformatics, University of Geneva, 1206, Genève, Switzerland}
    \item \textit{Italian Institute of Technology, Via Melen 83, 16152 Genoa, Italy}
    \item \textit{Department of Chemistry, University College London, London, WC1E 6BT, United Kingdom}
\end{enumerate}
\end{raggedright}


\captionsetup[figure]{labelsep=period, name=FIG SI.}

\setcounter{figure}{0}

\renewcommand{\thesection}{SI \arabic{section}}
\setcounter{section}{0}

\section{Differentiable k-nearest-neighbor selection}
\label{Appendix:DifferentiableKNN}
DeepLNE implements the continuous and differentiable k-NN in the spirit of Ref.~\onlinecite{Tobias2018}. First describing the 1-NN selection rule where we define the negative Euclidean distance $r$ between query point $q$ and the transformed training datapoints $\bm{x}_i$, to be $a=-r(q,\bm{x}_i)$. The expectation of $\bar{w}^1$ of the first index vector is given by
\begin{equation}
\label{eq:w}
\bar{w}_i^1= P [ w^1 = i | a^1,t] = \frac{e^{a^1/t}}{\sum_{i' \in I} e^{(a_i^1/t)}} ~,
\end{equation}
The distances are updated such that 
\begin{equation}
\label{eq:a}
\bar{a}^{j+1} = \bar{a}_i^j + log (1-\bar{w}_i^j ) ~,
\end{equation}
The updated distances are then used to calculate the expectation over the next index vector
\begin{equation}
\label{eq:wij}
\bar{w}_i^{j+1}= P [ w^{j+1} = i | a^{j+1},t] = \frac{e^{a^1{j+1}/t}}{\sum_{i' \in I} e^{(a_i^{j+1}/t)}} ~,
\end{equation}
The continuous nearest neighbors $\{\bm{x}^{\text{k-NN}}_1,...,\bm{x}^{\text{k-NN}}_k\}$ of $q$ using the expectations $\bar{w}^j$ follows as
\begin{equation}
\label{eq:xj}
\bm{x}^{\text{k-NN}}_j = \sum \bar{w}_i^j \bm{x}_i ~.
\end{equation}

A computational speed up during the training and application of DeepLNE can be achieved by decreasing the size of the weight matrix $\bar{w}$ used for k-NN. To this extent, one can use Farthest Point-Sampling (FPS) \cite{Goscinski2023} selecting a subset of the training datapoints only for the neighbors selection and the construction of $\bm{x}^\mathrm{k-NN}$.

\newpage
\section{Hyperparameter influence on neighborhood representation}
\begin{figure*}[h]
\includegraphics[width=\textwidth]{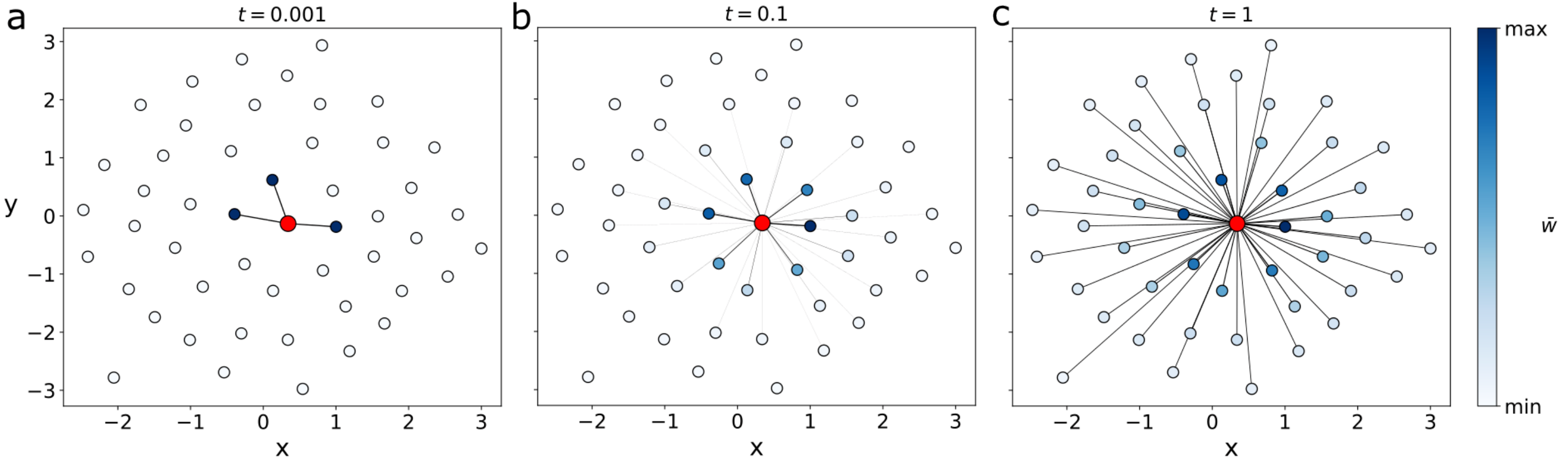}
\caption{\label{AppendixFig:KNN_t_dependency}
Depiction of the representation $\bm{x}^{\text{k-NN}}_i$ of a single datapoint $\bm{x}_i$ within its neighborhood for $k=3$, sampled from a bivariate normal distribution with zero mean and a covariance of 0.4 between dimensions. Various values of the hyperparameter $t$ are considered, and we visualize the weighted connections between datapoints using the selection matrix $\bar{w}$ (computed as described in Sec.~\ref{Appendix:DifferentiableKNN}).
(a) For $t=0.001$, exactly three neighbors contribute to $\bm{x}^{\text{k-NN}}_i$.
(b) With $t=0.1$, those three neighbors of $\bm{x}_i$ again exhibit high $\bar{w}$ values, however additional datapoints also contribute significantly.
(c) Increasing the hyperparameter to $t=1$ results in many datapoints contributing significantly to the neighborhood representation.
}
\end{figure*}

\newpage
\section{Particle in M\"uller-Brown-Potential}

\begin{figure*}[h]
\includegraphics[width=\textwidth]{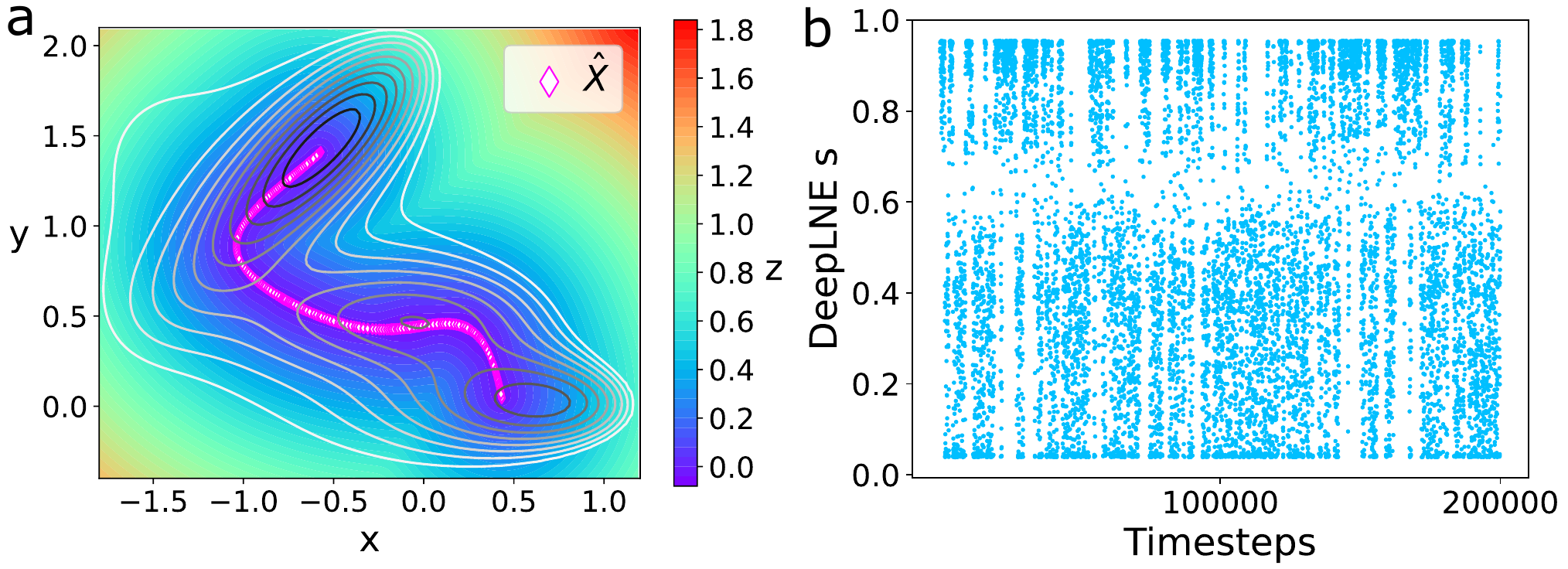}
\caption{\label{AppendixFig:MullerBrown}
Results of training DeepLNE for the M\"uller-Brown toymodel and using the path-like CV for biasing a new simulation on the fly. (a) We depict the decoded training datapoints $\bm{\hat{X}}$ of DeepLNE, which are used to compute the perpendicular distance the the path ($z$) for a chosen $\lambda=50$. We superimpose these data with the reference FES to show that the DeepLNE $z$ variable approximates well the path across the energetically lowest barrier. (b) Applying a harmonic constraint on $z$ during new enhanced sampling MD results in frequent transitions while sampling of the biased DeepLNE $s$ variable over time.}
\end{figure*}

\newpage
\section{Alanine dipeptide}

\begin{figure*}[h]
\includegraphics[width=\textwidth]{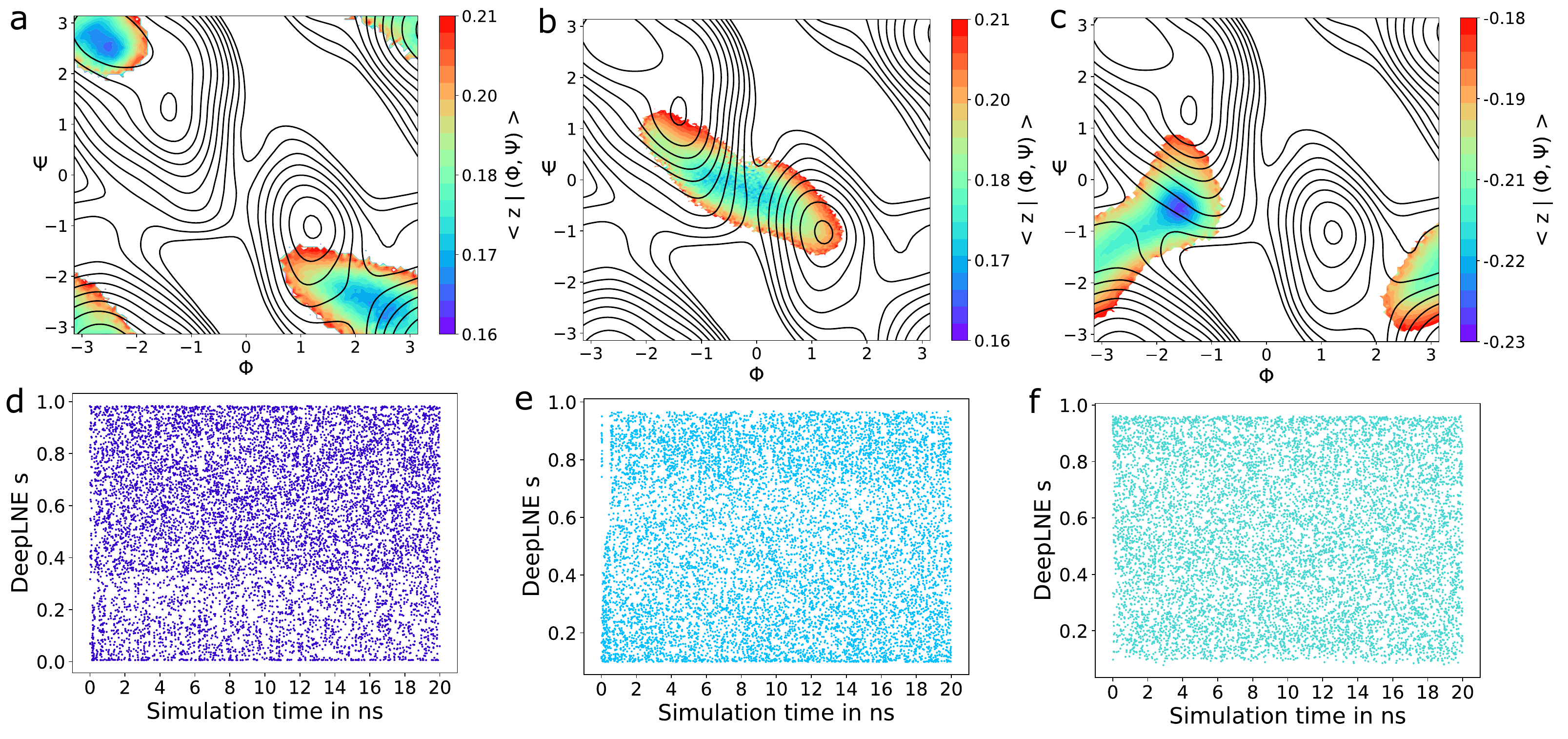}
\caption{\label{AppendixFig:Dialanine}
Since $\bm{\hat{X}}$ is of high dimensionality and can not be conveniently visualized we superimpose the expectation value $<z | (\phi,\psi)>$ of the DeepLNE CV based on the new alanine dipeptide simulations using OneOPES simulations with the reference FES of $\phi$ and $\psi$. In this way we try to visualize the perpendicular distance to the path-like CV automatically learned from the DeepLNE algorithm. (a) $<z | (\phi,\psi)>$ based on datapoints of the simulation along the path between $C_{7eq}$ and $C_{7ax}$ over a barrier at $\phi=2, \psi=-2.5$. (b) $<z | (\phi,\psi)>$ estimated for the transition between the states $C_{7eq}$ and state $C_{7ax}$, across a barrier located close to $\phi=0, \psi=0$. (c) $< $z$ | (\phi,\psi)>$  based on configurations of a path across the barrier close to $\phi=-2, \psi=-1$ beginning and ending in the state $C_{7eq}$. 
The resulting time-evolution of the biased DeepLNE CV $s$ when depositing a bias potential on it via OneOPES along the 3 different transition paths can be found in the row below (d), (e), (f), respectively.}
\end{figure*}

\newpage
\begin{figure*}[h]
\includegraphics[width=\textwidth]{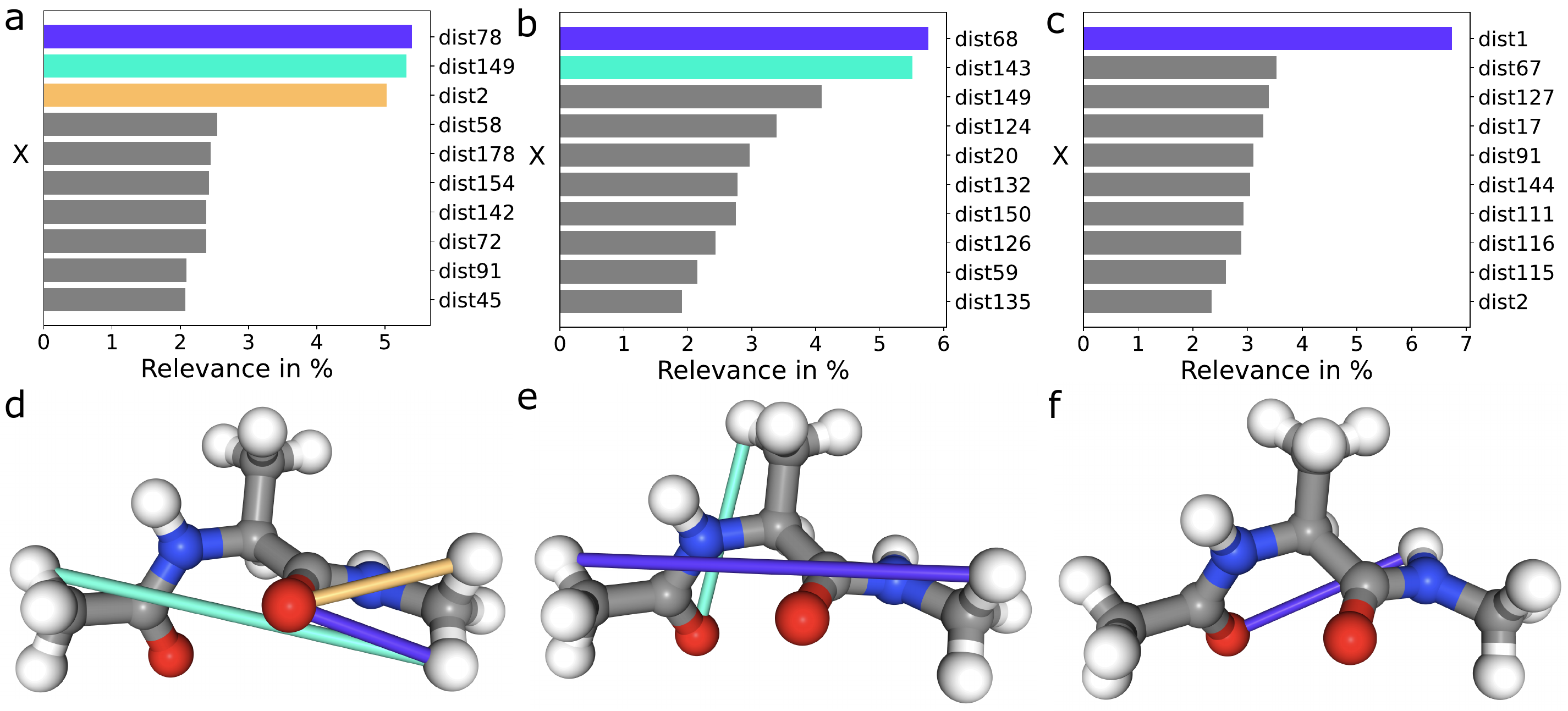}
\caption{\label{AppendixFig:Dialanine_FeatureRanking}
Input $X$ ranked based on magnitude of gradient of $s$ with respect to the $190$ input features $\bm{X}$. The 10 inputs with the highest gradients are shown for all 3 different transitions (a) path between $C_{7eq}$ and $C_{7ax}$ over the barrier at $\phi=2, \psi=-2.5$, (b) path between the states $C_{7eq}$ and state $C_{7ax}$ across the barrier at $\phi=0, \psi=0$, (c) path across the barrier at $\phi=-2, \psi=-1$ beginning and ending in the state $C_{7eq}$. The most relevant inputs colored in (a), (b) and (c) are visualized within the molecular structure of alanine dipeptide with the same color. For (d) and (e) $\phi$,$\psi$ both change significantly enough during the investigated transitions such that the DeepLNE model considers distances relevant, that are indeed sensitive to both angle changes. e) During the investigated transition $\psi$ changes most and accordingly the DeepLNE selects 1 relevant distance, that is sensitive to this dihedral change.}
\end{figure*}

\newpage
\section{RNA GAGA-tetraloop}

\begin{figure*}[h]
\includegraphics[width=\textwidth]{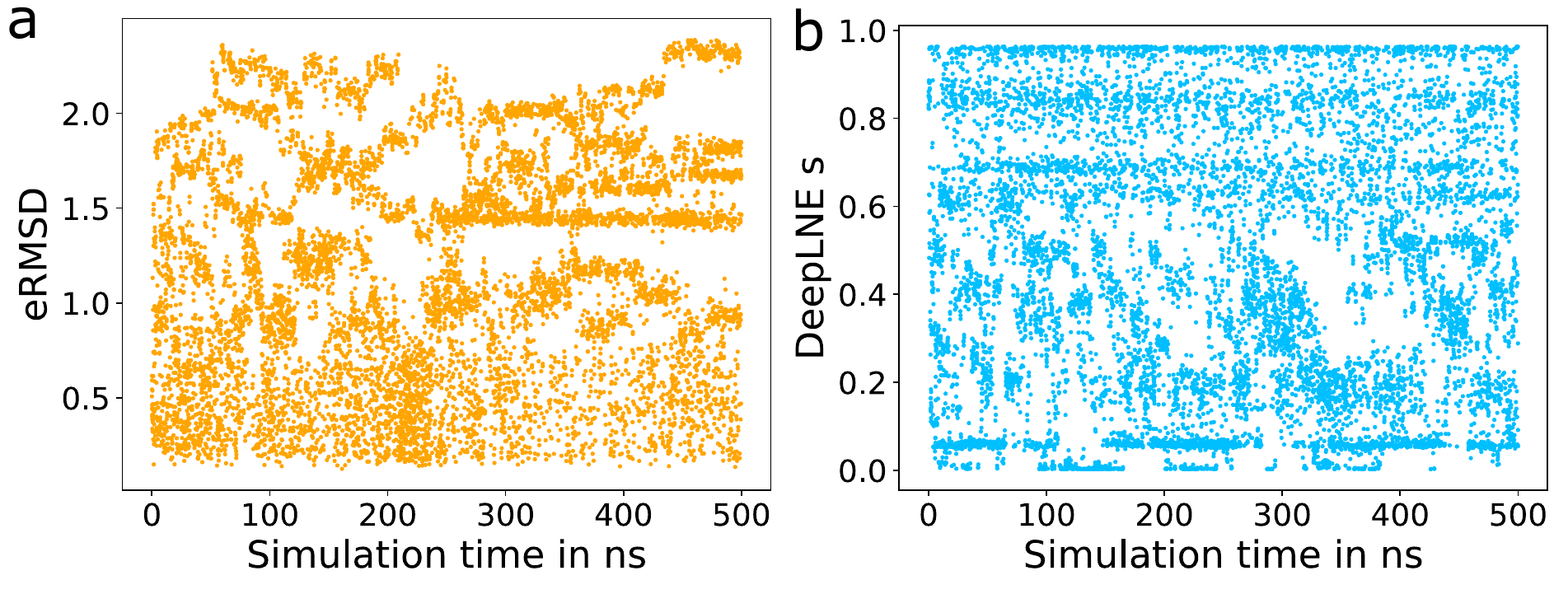}
\caption{\label{AppendixFig:RNA}
Comparison of timeseries of the biased CV for the OneOPES simulations of the RNA-Tetraloop.
(a) The eRMSD CV is not reliably exploring its entire range of values.
(b) Over the course of 500 $ns$ the biased DeepLNE $s$ variable explores the entire spectrum and frequent transitions between the natively folded state and misfolded states occur.}
\end{figure*}



\end{document}